\documentclass[11pt]{article}
\usepackage{amsmath}
\usepackage[dvips]{graphicx}
\begin{document}
\title{Comments on ``Perspectives on Galactic Dynamics via General Relativity''}
\author{D. Vogt\thanks{e-mail: danielvt@ifi.unicamp.br}\\
Instituto de F\'{\i}sica Gleb Wataghin, Universidade Estadual de Campinas\\
13083-970 Campinas, S.\ P., Brazil
\and
P. S. Letelier\thanks{e-mail: letelier@ime.unicamp.br}\\
Departamento de Matem\'{a}tica Aplicada-IMECC, Universidade Estadual\\
de Campinas 13083-970 Campinas, S.\ P., Brazil}
\maketitle
\begin{abstract}
In this comment we question some arguments presented in astro-ph/0512048 to 
refuse the presence of an singular mass surface layer. In particular, incorrect expressions 
are used for the disk's surface mass density. We also point out that
 the procedure of removing 
the descontinuity on the $z=0$ plane with a region of 
continuous density gradient generates 
other two regions of descontinuities with singular mass surface layers 
making the  model unrealistic. 
\end{abstract}
\section{Introduction}

In \cite{Cooperstock1} the authors make some comments about their previous work \cite{Cooperstock2} 
and also reply to certain issues that have been raised by some authors \cite{Korzynski,Vogt,Garfinkle}
 concerning the nature of the matter distribution and 
the asymptotic behaviour of the spacetime in their model. In particular, \cite{Korzynski}
pointed out that the use of a $|z|$ function in the solution of the field equations 
introduces an additional disk on the $z=0$ plane and later \cite{Vogt} showed that the disk 
was made of exotic matter.  Consequently, 
\cite{Cooperstock1} examine the question in several ways and present some arguments to dismiss the existence of a physical mass 
layer present on the $z=0$ plane. We would like to notice some inconsistencies in their paper. The first 
is related to the expressions used for the surface density of the singular disk, and is presented in 
Sec.\ \ref{sec_disk}. In Sec.\ \ref{sec_smooth} we question the smoothing
procedure discussed in  Section 4  of \cite{Cooperstock1}.  We find that one
cannot eliminate disk singularities, moreover one introduces new singularities
that make the model unrealistic.

\section{The disk's surface mass density} \label{sec_disk}

In \cite{Vogt} we calculate the distributional energy-momentum tensor $Q^a_b$ due to the introduction 
of an absolute value of $z$ in the metric functions. The resulting non-zero components read \cite{Vogt} 
\begin{subequations}
\begin{align}
Q^t_t &=\frac{1}{e^{\nu}}\left( \frac{NN_{,z}}{r^2}-\nu_{,z} \right) \mbox{,} \label{eq_Qtt} \\
Q^t_{\varphi} &=-\frac{N_{,z}}{e^{\nu}} \left( 1+\frac{N^2}{r^2} \right) \mbox{,} \\
Q^{\varphi}_t &=\frac{N_{,z}}{r^2e^{\nu}} \mbox{,} \\
Q^{\varphi}_{\varphi} &=-\frac{1}{e^{\nu}}\left( \frac{NN_{,z}}{r^2}+\nu_{,z} \right) \mbox{,} \label{eq_Qphiphi}
\end{align}
\end{subequations}
where all quantities are evaluated on $z=0$. Since $Q^a_b$ is 
non-diagonal, in order to have obtain the physical variables of the 
disk, we need to put the energy-momentum tensor in its canonical form.
 To do that  we  solve the eigenvalue problem: $Q^a_b\xi^b=\lambda\xi^a$. 
We thus find 
\begin{align}
& \lambda_{\pm}=\frac{T}{2}\pm \frac{\sqrt{D}}{2}, \quad \text{where} \\
& T=Q^t_t+Q^{\varphi}_{\varphi}, \qquad D=(Q^t_t-Q^{\varphi}_{\varphi})^2+4Q^t_{\varphi}Q^{\varphi}_t \mbox{,}
\end{align}
and using Eq.\ (\ref{eq_Qtt})--(\ref{eq_Qphiphi}) result in
\begin{equation}
T=-\frac{2\nu_{,z}}{e^{\nu}}, \qquad D=-\frac{4N_{,z}^2}{r^2e^{2\nu}} \mbox{.}
\end{equation}
As the discriminant is always negative, the eigenvalues and corresponding eigenfunctions
are complex conjugate. If $V_a$ and $W_a$ denote the timelike
 and spacelike real eigenvectors, 
the canonical form of the energy-momentum tensor is
\begin{equation} 
Q_{ab}=\sigma V_aV_b+p_{\varphi}W_aW_b+\kappa(V_aW_b+W_aV_b) \mbox{,}
\end{equation}
where $\sigma=T/2$ is the surface density, $p_{\varphi}=-T/2$ denote the azimuthal stresses 
and $\kappa=\sqrt{-D}/2$ is the heat flow in 
the azimuthal direction \cite{Moller}.  
Thus the expression for the surface density, to order $G^1$, is given by $\sigma=-\nu_{,z}$, or, using the 
relation $\nu_{,z}=-N_{,r}N_{,z}/r$ we get
\begin{equation} \label{eq_sigma}
\sigma=\frac{N_{,r}N_{,z}}{r} \mbox{.}
\end{equation}

On the other hand, \cite{Cooperstock1} take following expressions for the surface energy density (Equations (15) and (16) of
the paper)
\begin{equation} \label{eq_sigma2}
\sigma= \frac{NN_{,z}}{r^2}-\nu_{,z}=\frac{NN_{,z}}{r^2}+\frac{N_{,r}N_{,z}}{r} \mbox{,}
\end{equation}
which they integrate over the surface and compare with the volume integral of their continuous mass
density distributions. But Eq.\ (\ref{eq_sigma2}) is only the $Q^t_t$ component Eq.\ (\ref{eq_Qtt}) to order $G^1$.
Due to the non-diagonal form of the energy-momentum tensor, the $Q^t_t$ component solely does not determine 
the surface density, but there is also a contribution from the $Q^{\varphi}_{\varphi}$ component Eq.\ (\ref{eq_Qphiphi}) 
\emph{even to order} $G^1$. Thus, the correct expression for $\sigma$ that should be used is Eq.\ (\ref{eq_sigma}). 
The integration of this equation over the surface would result in half of the value for the mass derived from the volume 
integral of the continuous mass distribution, since the authors themselves comment on footnote 7 that the two terms
in Eq.\ (\ref{eq_sigma2}) contribute equally. 

It is important to stress that the physical variables of the mass layer on the $z=0$ plane
are not directly given by the principal diagonal terms of the distributional energy-momentum tensor,
since the non-diagonal terms are non-negligible. This is further an example of how the
non-linearity of General Relativity can manifest even at Newtonian level.

\section{The smoothing procedure} \label{sec_smooth}

Another approach used in \cite{Cooperstock1} to examine the presence of a singular mass layer on 
the symmetry plane was to smooth the solution over an interval that includes the $z=0$ plane. This was 
achieved by the choice of $\cosh (k_nz)$ functions to span the symmetry plane in the interval
$-z_0<z<z_0$. For $|z| \geq z_0$, the original solution with exponentials was used. This requires that
the functions $N$ \emph{and} also $N_{,z}$ match at $|z|=z_0$.
If this last condition is not satisfied new matter is added. We note that this
is also true for the Newtonian gravitational potential, discontinuity of the
first derivatives ``add matter''  whose density can be computed 
via Poisson's equation.
We argue that the above mentioned   matching   cannot be done without adding
new matter.

Let us take the function $N$ as follows:
\begin{subequations}
\begin{align}
N_1 &=-\sum_n C_{1n}k_ne^{k_nz}rJ_1(k_nr), \quad z\leq -z_0, \label{eq_N1}\\
N_2 &=-\sum_n C_{2n}k_n\cosh (k_nz)rJ_1(k_nr), \quad -z_0 <z<z_0, \label{eq_N2}\\
N_3 &=-\sum_n C_{3n}k_ne^{-k_nz}rJ_1(k_nr), \quad z\geq z_0, \label{eq_N3}
\end{align}
\end{subequations}
and impose the conditions $N_1(-z_0)=N_2(-z_0)$ and $N_{1,z}(-z_0)=N_{2,z}(-z_0)$. Using Eq.\ (\ref{eq_N1})--(\ref{eq_N2})
we obtain
\begin{align}
& \sum_n k_nrJ_1(k_nr) \left[ C_{2n}\cosh(k_nz_0)-C_{1n}e^{-k_nz_0} \right]=0, \label{eq_match1}\\
& \sum_n k_n^2rJ_1(k_nr) \left[ -C_{2n}\sinh(k_nz_0)-C_{1n}e^{-k_nz_0} \right]=0. \label{eq_match2}
\end{align}
Since the Bessel functions are linearly independent, the terms in brackets must vanish identically.
Subtracting Eq.\ (\ref{eq_match1}) from Eq.\ (\ref{eq_match2}) results in $C_{2n}e^{k_nz_0}=0$, which
is only satisfied in the real domain if all $C_{2n}=0$. Thus it is not possible to match $N$ and $N_{,z}$ 
simultaneously. 
\begin{figure}
\centering
\includegraphics[scale=0.6]{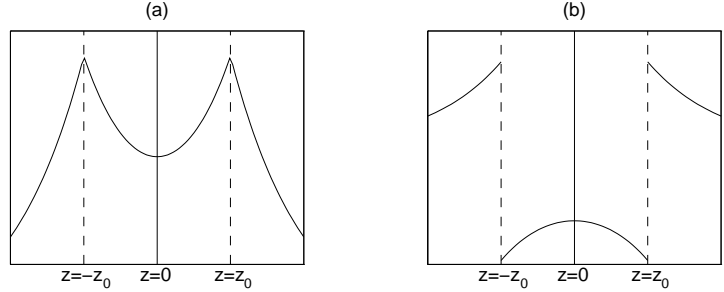}
\caption{Impossibility of matching the functions Eq.\ (\ref{eq_N1})--(\ref{eq_N3}) and their derivatives. 
(a) Continuity of $N$. (b) Continuity of $N_{,z}$.}\label{fig1}
\end{figure}
If we demand continuity of $N_1$, $N_2$ and $N_3$ at $|z|=z_0$ we obtain the following 
conditions: 
\begin{equation} \label{eq_cond}
C_{1n}=C_{3n}, \quad \text{and} \quad C_{2n}=C_{1n}\frac{e^{-k_nz_0}}{\cosh (k_nz_0)}. 
\end{equation}
On the other hand, demanding continuity of $N_{1,z}$, $N_{2,z}$ and $N_{3,z}$ at $|z|=z_0$ results in 
\begin{equation} 
C_{1n}=C_{3n}, \quad \text{and} \quad C_{2n}=-C_{1n}\frac{e^{-k_nz_0}}{\sinh (k_nz_0)}. 
\end{equation}
Fig.\ \ref{fig1}(a) sketches what happens if we impose continuity of 
the functions and Fig.\ \ref{fig1}(b) if we impose continuity of the derivatives. 
The only way the functions Eq.\ (\ref{eq_N1})--(\ref{eq_N3}) \emph{and} their 
derivatives could be matched is if an extra set of constants were inserted into Eq.\ (\ref{eq_N2})
\begin{equation}
N_2 =-\sum_n k_n(C_{2n}\cosh (k_nz)+C_{2n}^{\mathrm{extra}})rJ_1(k_nr)\mbox{,}
\end{equation}
but then this would not be a solution of 
\begin{equation}
N_{rr}+N_{zz}-\frac{N_r}{r}=0 \mbox{.}
\end{equation}
 
Assuming conditions Eq.\ (\ref{eq_cond}) hold, the jump of the derivatives of Eq.\ (\ref{eq_N1})--(\ref{eq_N3}) with respect to $z$ 
evaluated at $|z|=z_0$ are given by
\begin{equation}
N_{2,z}(-z_0)-N_{1,z}(-z_0)=N_{3,z}(z_0)-N_{2,z}(z_0)=\sum_n \frac{C_{1n}k_n^2rJ_1(k_nr)}{\cosh (k_nz_0)} \mbox{.}
\end{equation}
The same kind of descontinuities in the derivatives also appear when a $|z|$ is introduced in the solution. Thus they introduce 
additional layers of matter now located on the $z=\pm z_0$ planes that makes
the solution unrealistic. 
 
\bigskip
D. Vogt thanks CAPES for financial support. P. S. Letelier thanks CNPq and 
FAPESP for financial support.


\begin{thebibliography}{9}
\bibitem{Cooperstock1} F. I. Cooperstock and S. Tieu, preprint: astro-ph/0512048.
\bibitem{Cooperstock2} F. I. Cooperstock and S. Tieu, preprint: astro-ph/0507619.
\bibitem{Korzynski} M. Korzy\'nski, preprint: astro-ph/0508377. 
\bibitem{Vogt} D. Vogt and P. S. Letelier, preprint: astro-ph/0510750. 
\bibitem{Garfinkle} D. Garfinkle, preprint: gr-qc/0511082. 
\bibitem{Moller} C. M\o{}ller, \textit{The Theory of Relativity}, Oxford University Press, 1972.
\end{thebibliography}
\end{document}